\documentclass[twocolumn,english,aps,showpacs,showkeys]{revtex4}
\usepackage[T1]{fontenc}
\usepackage[latin9]{inputenc}
\usepackage{array}
\usepackage{textcomp}
\usepackage{relsize}
\usepackage{amsmath}
\usepackage{graphicx}

\makeatletter

\DeclareRobustCommand{\lyxmathsym}[1]{\ifmmode\begingroup\def\b@ld{bold}
  \def\rmorbf##1{\ifx\math@version\b@ld\textbf{##1}\else\textrm{##1}\fi}
  \mathchoice{\hbox{\rmorbf{#1}}}{\hbox{\rmorbf{#1}}}
  {\hbox{\smaller[2]\rmorbf{#1}}}{\hbox{\smaller[3]\rmorbf{#1}}}
  \endgroup\else#1\fi}

\providecommand{\tabularnewline}{\\}

\usepackage{longtable}
\usepackage{hyperref}
\def\R2Lurl#1#2{\mbox{\href{#1}{\tt #2}}}

\makeatother

\usepackage{babel}

\makeatother

\usepackage{babel}

\begin{document}

\title{Structural and magnetic properties of Co$_{2}$MnSi thin films}

\author{M. Belmeguenai$^{1}$\footnote{electronic adress: belmeguenai.mohamed@univ-paris13.fr}, F. Zighem$^{2}$, D. Faurie$^{1}$, S. M.
Chérif$^{1}$, P. Moch$^{1}$, K.Westerholt$^{3}$ and W. Seiler$^{4}$}

\affiliation{$^{1}$ \textit{\emph{LPMTM, Institut Galilée, UPR 9001 CNRS, Université
Paris 13,}} \textit{\emph{99 Avenue Jean-Baptiste Clément 93430 Villetaneuse,
France}}}

\affiliation{\emph{$^{2}$} \textit{\emph{IMCN, Université Catholique de Louvain,
Place Croix du Sud 1, 1348 LLN, Belgium}}}

\affiliation{\emph{$^{3}$} Experimentalphysik/Festkörperphysik, Ruhr-Universität
Bochum, 44780 Bochum, Germany}

\affiliation{\emph{$^{4}$} PIMM, ENSAM, 151 Boulevard de l'Hôpital, 75013 Paris,
France}
\begin{abstract}
Co$_{2}$MnSi (CMS) films of different thicknesses (20, 50 and 100
nm) were grown by radio frequency (RF) sputtering on a-plane sapphire
substrates. Our X-rays diffraction study shows that, in all the samples,
the cubic $<110>$ CSM axis is normal to the substrate and that there
exist well defined preferential in-plane orientations. Static and
dynamic magnetic properties were investigated using vibrating sample
magnetometry (VSM) and micro-strip line ferromagnetic resonance (MS-FMR),
respectively. From the resonance measurements versus the direction
and the amplitude of an applied magnetic field we derive most of the
magnetic parameters: magnetization, gyromagnetic factor, exchange
stiffness coefficient and magnetic anisotropy terms. The in-plane
anisotropy can be described as resulting from the superposition of
two terms showing a two-fold and a four-fold symmetry without necessarily
identical principal axes. The observed behavior of the hysteresis
loops is in agreement with this complex form of the in-plane anisotropy
\end{abstract}



\maketitle

\section{Introduction}

The strong spin polarization at the Fermi level of full Heusler alloys
and their high Curie temperature make of them potential candidates
for applications. Thus, these materials have been very recently inserted
in spintronics devices films made of Co$_{2}$MnSi full Heusler, which
has a Curie temperature of 985 K {[}1,2{]}, in order to be used in
magnetic tunnel junctions (MTJs) containing one or two Co$_{2}$MnSi
electrodes and different barriers {[}3-8{]}. Heusler alloys were used
as MTJ electrodes with an amorphous Al-oxide barrier {[}9-12{]} allowing
for a tunnel magnetoresistance (TMR) ratio of $159\%$ at 2 K in a
Co$_{2}$MnSi/Al-O/CoFe structure {[}6{]}. A large TMR ratio, up to
$753\%$ at 2 K and of $217\%$ at room temperature, was obtained
with Co$_{2}$MnSi used in TMJs with MgO barriers {[}13{]}.

Despite this intense research activity on Heusler alloys, which is
mainly focused on the way to improve the tunnel magnetoresistance,
the static and dynamic magnetic properties of such alloys remain less
explored {[}14-17{]}. The dynamics of these materials within the 1-10
GHz frequency range, which determines the high-speed response, is
a key for their future technologic applications, especially in view
of increasing data rates in magnetic storage devices. Moreover, the
exchange stiffness constant $A_{ex}$ which describes the strength
of the exchange interaction inside Co$_{2}$MnSi films is an important
parameter from both fundamental and application points of view. Therefore,
the aim of this paper is to investigate the static and dynamic magnetic
properties of Co$_{2}$MnSi thin films and their relations with their
structure.

The paper is organized as follows: in section II we briefly present
the preparation of the samples and their structural properties investigated
by X-rays diffraction (XRD). Section III exposes their main static
magnetic characteristics derived from our vibrating sample magnetometry
(VSM) measurements. Section IV presents and discusses our dynamic
measurements performed with the help of micro-strip ferromagnetic
resonance (MS-FMR). In section V, conclusions are drawn.

\section{Sample preparation and structural properties}

The Co$_{2}$MnSi thin films (20, 50 and 100 nm in thickness) were
deposited on a-plane sapphire substrate by UHV-magnetron rf-sputtering
using pure Ar at a pressure of $5\times10^{-3}$ mbar as sputter gas.
The base pressure of the sputtering system was $2\times10^{-9}$ mbar,
the sputtering rate was 0.03 nm.s$^{-1}$ for Co$_{2}$MnSi and 0.025
nm.s$^{-1}$ for the previously deposited vanadium seed underlayer.
During the deposition the temperature of the substrates was kept constant
at $470\lyxmathsym{\textdegree}$C. Heusler alloy targets with a diameter
of 10 cm were cut from single phase, stoichiometric ingots prepared
by high frequency melting of the components in high purity graphite
crucibles. The Al$_{2}$O$_{3}$ sapphire a-plane substrate was, as
mentioned above, preliminarily covered with a 5 nm thick vanadium-seed-layer
in order to induce a high quality $(110)$ growth of the Heusler compound.
After cooling them down to room temperature all the films were subsequently
covered by a 5 nm thick gold layer protecting them against oxidation.\\

\begin{figure}
\includegraphics[bb=30bp 130bp 350bp 580bp,clip,width=8.5cm]{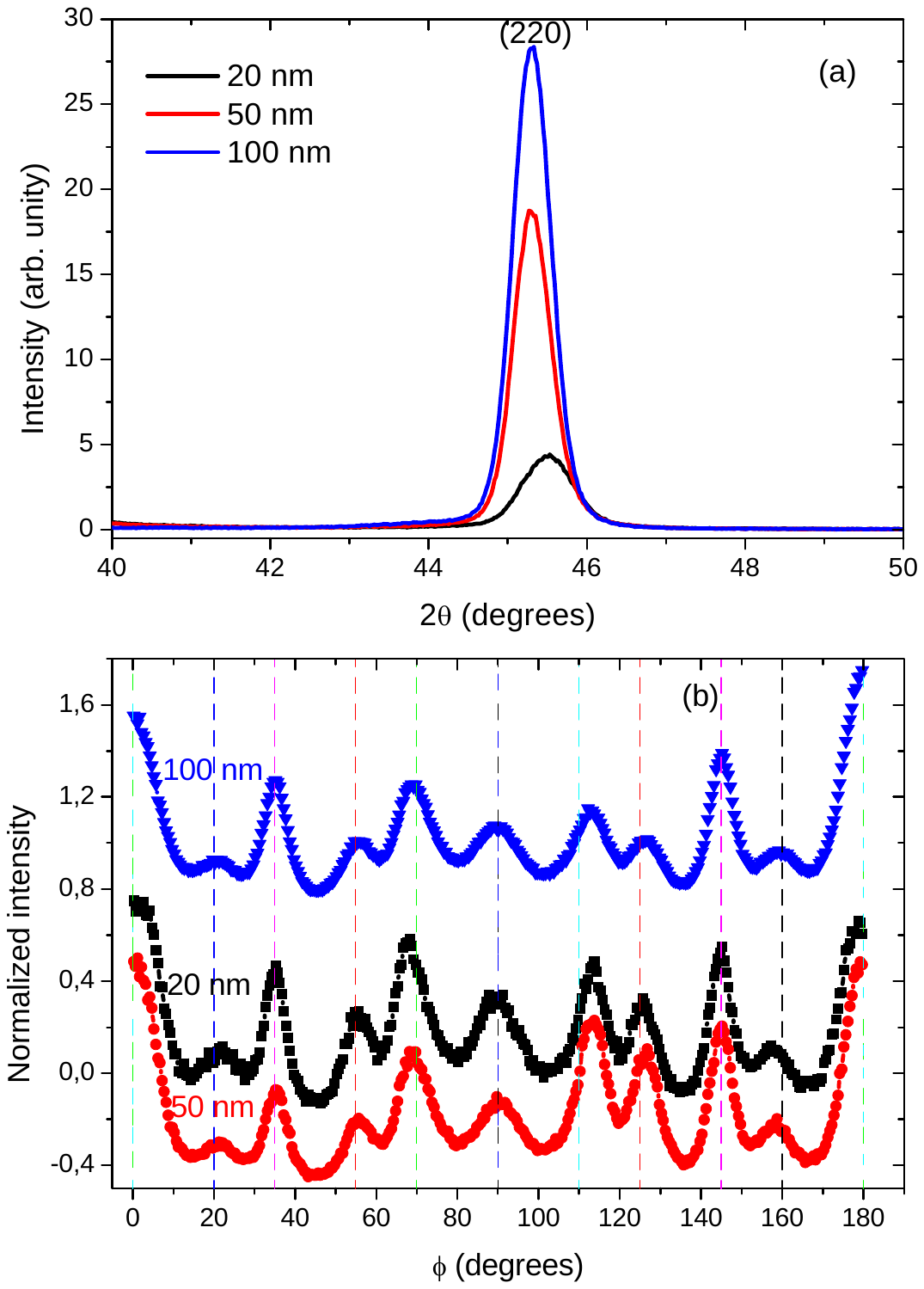}

\caption{(Color online) (a) X-ray Bragg scan using the Cu K$_{\alpha1}$ radiation
for the 100 nm thick Co$_{2}$MnSi thin film. (b) Angular variations
of the intensity around $60^{\circ}$ for different Co$_{2}$MnSi
thin films. Graphs are shifted vertically with a respect to that of
20 nm thick samples for clearness. The vertical color and dashed lines
refer to the expected positions of the diffraction peak relative to
the different variants belonging to the observed 4 families.}

\end{figure}

XRD measurements were performed using a Four-Circle diffractometer
in Bragg-Brentano geometry in order to obtain $\theta-2\theta$ patterns
and $\phi$-scans, operating at 40 kV and 40 mA, and using a Cu X-rays
source ($\lambda=0.1518$ nm). In all the films, the $\theta-2\theta$
pattern (Figure 1a) indicates that a $<110>$-type cubic axis is normal
to the sample plane. The Co$_{2}$MnSi deduced cubic lattice constant
(e.g.: $a\lyxmathsym{\lyxmathsym{ }}=5.658$ Å for the 100 nm thich
sample) is in good agreement with the previously published value (5.654
Å) {[}2{]}. The samples behave as ${110}$ fiber textures containing
well defined zones showing significantly higher intensities, as shown
in Figure 1b, which represents $\phi$-scans at $\psi=\lyxmathsym{\lyxmathsym{ }}60^{\circ}$
(here, $\psi$ is the declination angle between the scattering vector
$q$ and the direction normal to the film, $\phi$ is the rotational
angle around this direction). The regions related to the observed
maxima correspond to orientation variants which can be grouped into
four families. Two of them were observed in a previous study {[}18,
19{]} concerning thin films of a neighboring Heusler compound, Co$_{2}$MnGe,
prepared using an identical protocol: in the first one the threefold
or axis is oriented along the c rhombohedral direction of the sapphire
substrate, thus defining two kinds of distinct domains respectively
characterized by their {[}001{]} axis inclined at $+54.5^{\circ}$
or at $-54.5^{\circ}$ with respect to this $c$ orientation. The
second family is rotated by $90^{\circ}$ from the first one and also
contains two variants. In addition, we observe a third family consisting
in domains with their $[001]$ axis along c and a fourth one with
their $[001]$ axis normal to c. This distribution into 6 kinds of
domains does not appreciably vary from sample to sample, as shown
on Figure 1b. This difference in the structure between the Co$_{2}$MnSi
and Co$_{2}$MnGe films probably arises from the variation of the
lattice mismatch.

\section{Magnetic properties}

\subsection{Static magnetic measurements}

\begin{figure}
\includegraphics[bb=30bp 420bp 250bp 842bp,clip,width=8.5cm]{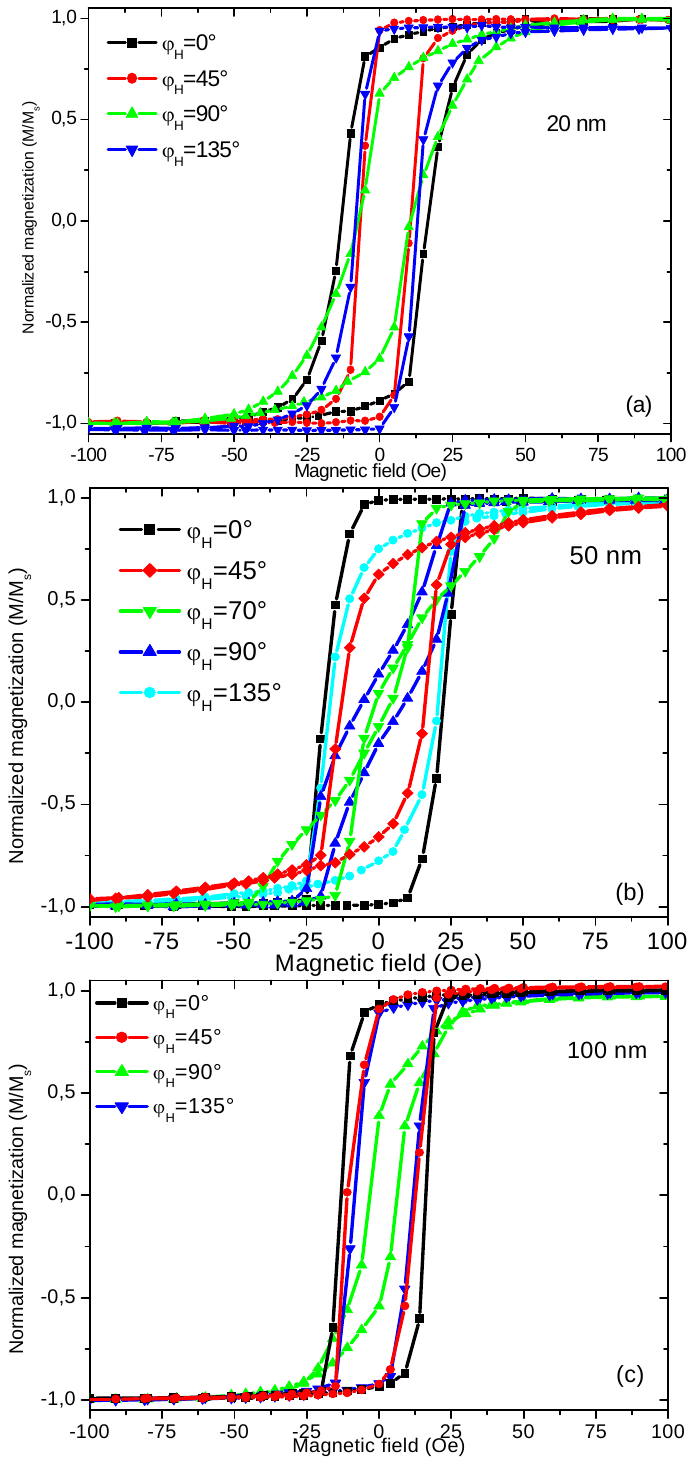}

\caption{(Color online) VSM magnetization loops of the (a) 20 nm thick, (b)
50 nm thick and (c) 100 nm thick Co$_{2}$MnSi samples. The magnetic
field is applied parallel to the film surface, at various angles ($\varphi{}_{\mathit{H}}$)
with the $c$-axis of the sapphire substrate.}

\label{Fig1:images-3}
\end{figure}

For all the samples the hysteresis curves were studied at room temperature
with an in-plane magnetic field $H$ applied along various orientations,
as shown in Figure 2 ($\varphi_{H}$ is the angle between $H$ and
the $c$-axis of the substrate). The variations of the reduced remanent
magnetization ($M_{r}/M_{s}$) as function of $\varphi_{H}$ are also
depicted in Figure 4, in view of comparison with other results discussed
in the next section. For any given direction of the applied field
the shape of the hysteresis loop is sample dependent, suggesting significant
differences in the amplitudes and in the principal directions describing
the in-plane anisotropy. Let us briefly discuss the case of the 50
nm film: as shown on Figure 2b, when $H$ lies along c ($\varphi_{H}\lyxmathsym{\lyxmathsym{ }}=\lyxmathsym{\lyxmathsym{ }}0^{\circ}$)
a typical easy axis square-shaped loop is observed, with a full normalized
remanence ($M_{r}/M_{s}=0.99$) and a coercive field of 20 Oe. As
$\varphi_{H}$ increases $M_{r}/M_{s}$ decreases and the hysteresis
curve tends to transform into a hard axis loop. When $\varphi_{H}$
reaches $70^{\circ}$ its shape becomes more complicated: it consists
into three smaller loops. A further increase of jH restores an almost
rectangular shape. Notice the observed difference between orthogonal
directions ($\varphi_{H}=45^{\circ}$ vs $\varphi_{H}=135^{\circ}$
or: $\varphi_{H}=-45^{\circ}$ vs $\varphi_{H}=-135^{\circ}$) which
prevents a simple interpretation based on a four-fold in-plane anisotropy.
Our data qualitatively agree with a description of the in-plane anisotropy
in terms of an addition of four-fold and two-fold contributions with
slightly misaligned easy axes. \\

Figures 2a and 2c show a series of hysteresis loops related to the
other films: here again, they qualitatively agree with the above description,
but with different relative contributions and orientations of the
two-fold and of the four-fold anisotropy terms. At evidence, a quantitative
estimation of the pertinent in-plane anisotropy terms monitoring the
dynamic properties presented in the next section cannot be derived
assuming hysteresis behaviour based on the coherent rotation model.
However, as discussed in the following, this model provides a satisfactory
account of the angular variation of $M_{r}/M_{s}$.

\begin{figure}
\includegraphics[bb=50bp 410bp 310bp 590bp,clip,width=8.5cm]{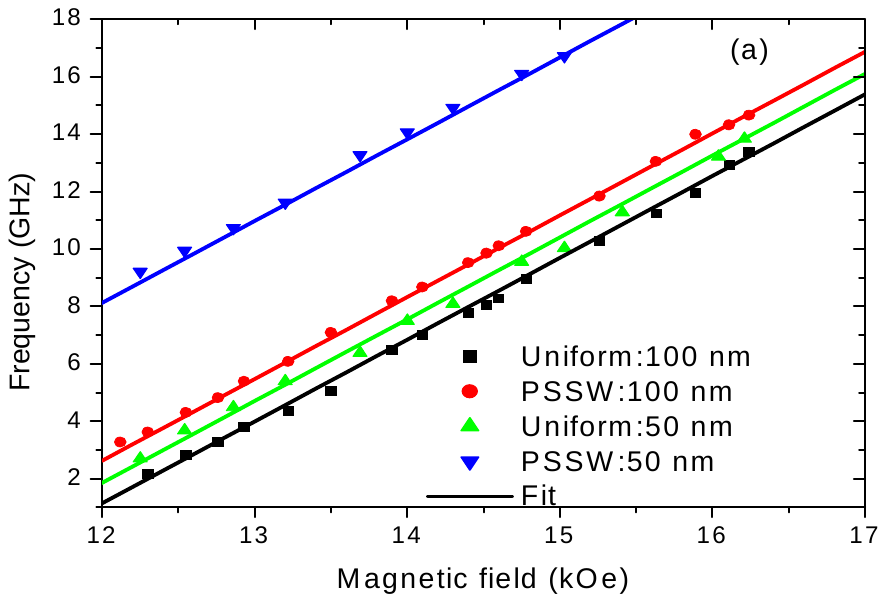}

\caption{(Color online) Field-dependence of the resonance frequency of the
uniform precession and of the first perpendicular standing spin wave
excited (PSSW) mode of 50 nm and 100 nm thick films. The magnetic
field is applied perpendicular to the film plane. The fits are obtained
using equation (3) with the parameters indicated in the Table I.}

\label{Fig1:images-3-1}
\end{figure}

\begin{table*}
\begin{tabular}{>{\centering}p{2cm}>{\centering}p{2cm}>{\centering}p{1cm}>{\centering}p{1cm}>{\centering}p{1cm}>{\centering}p{1cm}>{\centering}p{1cm}>{\centering}p{1cm}>{\centering}p{1cm}>{\centering}p{1cm}>{\centering}p{1cm}>{\centering}p{1cm}}
\multicolumn{12}{c}{}\tabularnewline
\hline
\hline
$d$

(nm)  & $D$

(\textmu{}erg.cm$^{-1}$)  & \multicolumn{2}{>{\centering}p{2cm}}{$4\pi M{}_{\mathit{eff}}$

(kOe) } & \multicolumn{2}{>{\centering}p{1.5cm}}{$H{}_{\mathit{u}}$

(Oe)} & \multicolumn{2}{>{\centering}p{2cm}}{$H{}_{\mathit{4}}$

(Oe)} & \multicolumn{2}{>{\centering}p{2cm}}{$\varphi{}_{\mathit{u}}$

(deg.)} & \multicolumn{2}{>{\centering}p{2cm}}{$\varphi{}_{4}$

(deg.)}\tabularnewline
\hline
\hline
20  &  & \multicolumn{2}{c}{11 } & \multicolumn{2}{>{\centering}p{2cm}}{20} & \multicolumn{2}{>{\centering}p{2cm}}{72} & \multicolumn{2}{>{\centering}p{2cm}}{0} & \multicolumn{2}{>{\centering}p{2cm}}{45}\tabularnewline
50  & 2.87  & \multicolumn{2}{>{\centering}p{2cm}}{11.35} & \multicolumn{2}{>{\centering}p{2cm}}{26 } & \multicolumn{2}{>{\centering}p{2cm}}{88} & \multicolumn{2}{>{\centering}p{2cm}}{-13} & \multicolumn{2}{>{\centering}p{2cm}}{0}\tabularnewline
100  & 2.56 & \multicolumn{2}{>{\centering}p{2cm}}{11.6} & \multicolumn{2}{>{\centering}p{2cm}}{16} & \multicolumn{2}{>{\centering}p{2cm}}{48} & \multicolumn{2}{>{\centering}p{2cm}}{0} & \multicolumn{2}{>{\centering}p{2cm}}{45}\tabularnewline
\hline
\hline
 &  &  &  &  &  &  &  &  &  &  & \tabularnewline
\end{tabular}

\caption{Magnetic parameters obtained from the best fits to our experimental
results. $\varphi{}_{\mathit{u}}$ and $\varphi{}_{\mathit{4}}$ are
the angles of the in-plane uniaxial and of the four-fold anisotropy
easy axes, respectively.}

\end{table*}

\subsection{Dynamic magnetic properties}

The dynamic magnetic properties were scrutinized using with the help
of a previously described MS-FMR {[}18, 19{]} setup. The resonance
frequencies are obtained from a fit assuming a lorentzian derivative
shape of the recorded spectra. As in ref. {[}18{]}, we assume a magnetic
energy density which, in addition to Zeeman, demagnetizing and exchange
terms, is characterized by the following anisotropy contribution:

$E_{anis.}=K_{\perp}\sin^{2}\theta_{M}-\frac{1}{2}(1+cos(2(\varphi_{M}-\varphi_{u}))K_{u}\sin^{2}\theta_{M}-\frac{1}{8}(3+\cos4(\varphi_{M}-\varphi_{4}))K_{4}\sin^{4}\theta_{M}$
(1)\\

In the above expression, $\theta{}_{\mathit{M}}$ and $\varphi{}_{\mathit{M}}$
respectively represent the out-of-plane and the in-plane (referring
to the $c$-axis of the substrate) angles defining the direction of
the magnetization $M_{s}$ ; $\varphi_{u}$ and $\varphi_{4}$ stand
for the angles of the uniaxial axis and of the easy fourfold axis,
respectively, with this $c$-axis. With these definitions $K_{u}$
and $K_{4}$ are necessarily positive. As in ref. {[}18{]}, it is
convenient to introduce the effective magnetization $4\pi M_{eff}=4\pi M_{s}-2K_{\perp}/M_{s}$,
the uniaxial in-plane anisotropy field $H_{u}\lyxmathsym{\lyxmathsym{ }}=\lyxmathsym{\lyxmathsym{ }}2K_{u}/M_{s}$
and the fourfold in-plane anisotropy field $H_{4}\lyxmathsym{\lyxmathsym{ }}=\lyxmathsym{\lyxmathsym{ }}4K_{4}/M_{s}$.\\

For an in-plane applied magnetic field $H$, the studied model provides
the following expression of the frequencies of the experimentally
observable magnetic modes:

$F_{n}\text{\texttwosuperior}=\left(\frac{\gamma}{2\pi}\right)^{2}(H\cos(\varphi_{H}-\varphi_{M})+\frac{H{}_{4}}{2}\cos4(\varphi_{M}-\varphi_{4})+H_{u}\cos2(\varphi_{M}-\varphi_{u})+\frac{2A_{ex.}}{M_{s}}\left(\frac{n\pi}{d}\right)^{2})\times(H\cos(\varphi_{H}-\varphi_{M})+4\pi M_{eff}+\frac{H_{4}}{2}(3+\cos4(\varphi_{M}-\varphi_{4}))+\frac{H_{u}}{2}(1+\cos2(\varphi_{M}-\varphi_{u}))+\frac{2A_{ex.}}{M_{s}}(\frac{n\pi}{d})^{2})\,\,\,\,\,\,\,\,\,(2)$\\

For $H$ normal to the sample plane, large magnetic fields are applied
(enough to allow for $M_{s}\lyxmathsym{\lyxmathsym{ }}\parallel\lyxmathsym{\lyxmathsym{ }}H$)
and thus the magnetic in-plane anisotropies can be neglected. Therefore,
the frequency linearly varies versus H as:

$F_{\perp}=\frac{\gamma}{2\pi}(H-4\pi M_{eff}+\frac{2A_{ex.}}{M_{s}}\left(\frac{n\pi}{d}\right)^{2})$
(3)\\

In the above expressions the gyromagnetic factor \textit{$\gamma$}
is related to the effective Lande factor g through: \textit{$(\gamma/2\pi)=g\times1.397\times10^{6}$}
s$^{-1}$.Oe$^{-1}$. The uniform mode corresponds to $n\lyxmathsym{\lyxmathsym{ }}=\lyxmathsym{\lyxmathsym{ }}0$.
The other modes to be considered (perpendicular standing spin waves:
PSSW) are connected to integer values of $n$: their frequencies depend
upon the exchange stiffness constant $A_{ex}$ and upon the film thickness
$d$.

For all the films the magnetic parameters at room temperature were
derived from MS-FMR measurements.

\subsubsection{Gyromagnetic g-factor and spin wave stiffness}

In perpendicular configuration the MS-FMR technique allows for deriving
the values of $g$ and of $4\pi M_{eff}$ from the variation of the
resonance frequency versus the magnitude of the applied field using
equation (3). The MS-FMR field-dependences of the resonance frequencies
of the uniform and of the PSSW modes are shown on Figure 3 for the
50 nm and the 100 nm thick samples. The frequencies vary linearly
with $H$. The PSSW mode shows a frequency higher than the uniform
one, by an amount independent of $H$, as expected from the studied
model. The derived value of $g$ is independent of the sample: $g=2.04$.
It is in good agreement with previous determinations {[}15{]}. The
effective demagnetizing field $4\pi M_{eff}$ slightly increases versus
the sample thickness but remains close to the saturation magnetization
(12200 Oe) given by Hamrle et al. {[}20{]}: it is reported in Table
I. The best fits for the observed PSSW are obtained using $A_{ex}=2.87\times10^{-6}$
erg.cm$^{-1}$ and $A_{ex}=2.56\times10^{-6}$ erg.cm$^{-1}$ for
the 50 nm and the 100 nm thick films respectively, in good agreement
with the result published by Hamrle et al.($A_{ex}=2.35\times10^{-6}$
erg.cm$^{-1}$) {[}20{]}.

\begin{figure}
\includegraphics[bb=30bp 15bp 290bp 585bp,clip,width=8.5cm]{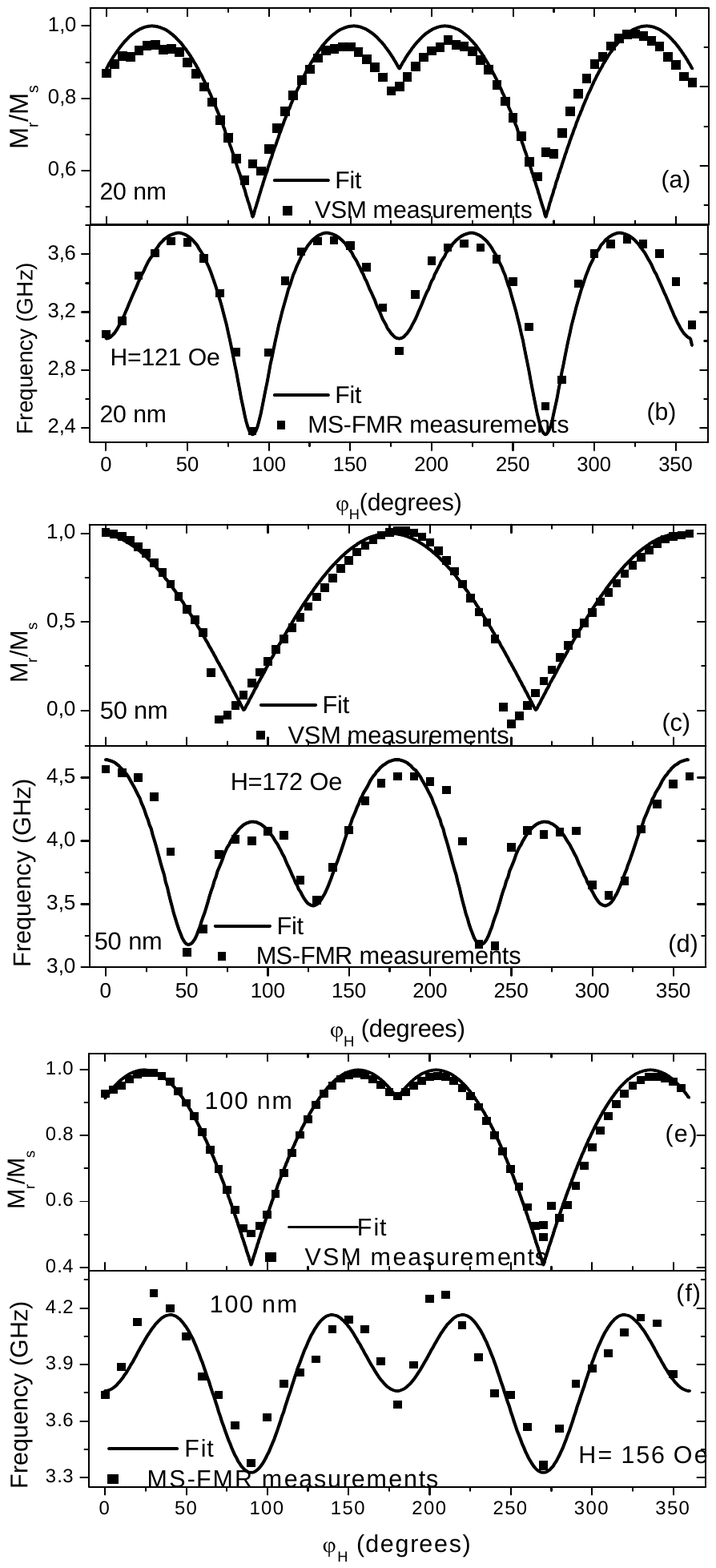}

\caption{Reduced remanent magnetization of the (a) 20 nm, of the (c) 50 nm
and of (e) 100 nm thick Co$_{2}$MnSi films. The full lines are obtained
from the energy minimization using the parameters reported in Table
I. (b), (d) and (f) show the compared in-plane angular-dependences
of the resonance frequency of the uniform modes. The fit is obtained
using equation (2) with the parameters indicated in Table I.}

\label{Fig1:images-3-2}
\end{figure}

\subsubsection{In-plane anisotropies}

Figures 4b, 4d and 4f illustrate the experimental in-plane angular
dependence of the frequency of the uniform mode in the 20, 50 and
100 nm thick films, compared to the obtained fits using equation (2).
For all the samples, the obtained values of the magnetic parameters
corresponding to the best fits are reported in Table I.

In all the investigated films the c-axis of the substrate coincides
with a principal direction of the four-fold magnetic anisotropy: it
defines a hard axis ($\varphi_{4}=45^{\circ}$) except, surprisingly,
in the 50 nm sample for which it defines an easy axis ($\varphi_{4}=0^{\circ}$).
The directions of the principal axes of the two-fold anisotropy are
sample dependent. The observed variations of the in-plane magnetic
anisotropy are not clearly related either to the thickness or to the
crystallographic texture (which does not significantly change). During
the preparation of the films uncontrolled parameters presumably induce
different stress conditions giving rise to changes in the magnetic
anisotropy. Finally, it is interesting to notice that the set of in-plane
magnetic anisotropy parameters deduced from the in-plane angular dependence
of the magnetic resonance for the 20 nm, 50 nm and 100 nm thick samples
allows for a good fit of the angular variation of the normalized static
remanence calculated with the help of the coherent rotation model,
as shown in Figures 4. This variation only depends on $\varphi_{u}$,
$\varphi_{4}$ and $H_{u}/H_{4}$.

It results from the present study and from our previous ones {[}18,
19{]} that the growth of Heusler thin films on sapphire substrates
is complex and depends on the conditions of the sample preparation.
Therefore the magnetic properties cannot be tuned rigorously.

\section{Conclusion}

Co$_{2}$MnSi films of thicknesses varying from 20, 50 and 100 nm
were prepared by sputtering on a $a$-plane sapphire substrate. They
show practically identical crystallographic textures, as revealed
by our X-rays diffraction studies, with a cubic $[110]$ axis normal
to the film plane and with a well defined manifold of in-plane orientations
referring to the $c$-axis of the substrate. The micro-strip ferromagnetic
resonance gives access to effective g factors which do not differ
from each other and to effective demagnetizing fields which are close
to the magnetization at saturation and which slightly increase with
the sample thickness. In addition, this technique allowed us for evaluating
the exchange stiffness constant which, as expected, was found not
to significantly depend on the studied sample. The in-plane anisotropy
was investigated through the study of the dependence of the resonance
frequency versus the orientation of an in-plane applied magnetic field:
it presents two contributions, showing a four-fold and a two-fold
axial symmetry, respectively. A principal four-fold axis parallel
to $c$ is always found, but it can be the easy or the hard axis.
Depending on the studied film the two-fold axes can be misaligned
with the four-fold ones. The angular dependence of the remanent normalized
magnetization, studied by VSM and analyzed within the frame of a coherent
rotation model, is in agreement with these conclusions. The general
magnetic behavior is similar to the previously published one observed
in Heusler Co$_{2}$MnGe films {[}18, 19{]}. Apparently, there is
no simple relation between the observed dispersion of the in-plane
anisotropy parameters and the thickness or the crystallographic texture
of the samples.

\end{document}